\documentclass{moriond}
\usepackage{amsmath}
\usepackage{multirow}
\usepackage{multicol}

\newcommand{\abs}[1]{\left\lvert#1\right\rvert}

\newcommand{\Tr}[1]{\text{Tr}\left[#1\right]}

\def\be{\begin{equation}}
\def\ee{\end{equation}}
\def\bea{\begin{eqnarray}}
\def\eea{\end{eqnarray}}

\newcommand{\Photo}{\includegraphics[height=35mm]{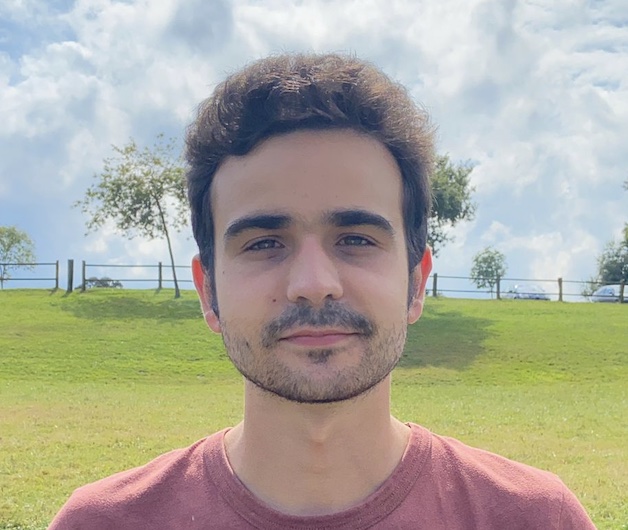}}

\begin{document}
\vspace*{4cm}
\title{New global bounds on heavy neutrino mixing}

\author{ Daniel Naredo }

\address{Departamento de F\'{\i}sica Te\'orica and Instituto de F\'{\i}sica Te\'orica UAM/CSIC,\\
Universidad Aut\'onoma de Madrid, Cantoblanco, 28049 Madrid, Spain
}

\maketitle\abstracts{
We present an updated and improved global fit analysis of precision and flavor observables so as to derive constraints on deviations from unitarity in the leptonic mixing matrix. We focus on the case in which these deviations are sourced by heavy neutrinos, such that our results can also be interpreted as bounds on their mixing with the active flavors. Our analysis is motivated by new results on key observables, such as $M_W$, $|V_{ud}|$ and $\Gamma_{\text{inv}}$.
}

\section{Introduction}
The addition of heavy right-handed neutrinos to the Standard Model (SM) of particle physics is arguably one of the simplest ways of addressing several of the open problems in particle physics. If their mass lies above the electroweak scale, they become too heavy to be produced in experiments, such as colliders, thus making it not possible to directly probe them. However, their presence induces deviations from unitarity in the leptonic mixing matrix~\cite{Broncano:2002rw}, which can be probed by means electroweak precision observables (EWPO), lepton flavor universality (LFU) ratios, measurements of the Cabibbo-Kobayashi-Maskawa (CKM) matrix and charged lepton flavor violating (cLFV) processes.

Heavy neutrinos above the electroweak scale can be safely integrated out, giving rise to several effective operators. At the lowest order, they generate the very well-known dim-5 Weinberg operator responsible for neutrino masses, whose coefficient is given by the seesaw relation
\begin{equation}
    m_\nu=-\Theta M_M \Theta^T\,,
    \label{mnu}
\end{equation}
where $\Theta$ is the mixing matrix between heavy and SM light neutrinos, given by the ratio between the Dirac and Majorana mass matrices $\Theta\equiv m_D M_M^{-1}$. At dim-6, the only operator generated at tree level induces the aforementioned deviations from unitarity in the leptonic mixing matrix $N$ which, in full generality, can be parametrised as
\begin{equation}
    N=(1-\eta)\hspace{0.06cm}U\,,
\end{equation}
where $\eta$ is a hermitian matrix and $U$ is the unitary matrix diagonalising the Weinberg operator. This is a convenient parametrisation, as the non-unitarity effects are directly encoded in $\eta$, whose structure will be model dependent. In the particular context of heavy right-handed neutrinos:
\begin{equation}
    \eta=\dfrac{1}{2}\Theta\Theta^\dagger\,.
\end{equation}

This implies that, in this setup, $\eta$ is a positive definite matrix, which entails the following constraints on its entries:
\begin{equation}
    \eta_{\alpha\alpha}\geq0\,,\hspace{1cm}\left\lvert \eta_{\alpha\beta}\right\rvert\leq \sqrt{\eta_{\alpha\alpha}\eta_{\beta\beta}}\hspace{0.25cm}
    \text{(Schwarz inequality)}\,\hspace{1cm}\alpha,\beta=e,\mu,\tau
    \label{etapositive}
\end{equation}

Here we present an updated global fit analysis of several precision and flavor observables so as to derive constraints on non-unitarity deviations sourced by heavy neutrinos, which directly translate into bounds on their mixing with the active neutrino flavors. These bounds~\cite{Blennow:2023mqx} are the strongest for heavy neutrinos above the electroweak scale.

\section{Observables and Analysis}
We build our test-statistics from the following sets of observables:
\begin{itemize}
    \item[$\bullet$] Four determinations of the $W$-boson mass from LEP, Tevatron, ATLAS and LHCb. We do not include the anomalous CDF-II measurement, as we find it is in strong tension with other EWPO~\cite{Blennow:2023mqx}.
    \item[$\bullet$] Two determinations of the weak effective angle by Tevatron and LHC.
    \item[$\bullet$] Five Z-pole observables from LEP and a determination of the Z-boson invisible width by CMS.
    \item[$\bullet$] Five weak decay ratios constraining lepton flavor universality of weak interactions.
    \item[$\bullet$] Ten weak decays constraining CKM unitarity.
    \item[$\bullet$] Bounds on cLFV processes.
\end{itemize}

All of the above observables constrain the diagonal entries of $\eta$, with the exception of cLFV processes, which can directly constrain the off-diagonal entries. Alternatively, the Schwarz inequality (see Eq.~\ref{etapositive}) allows to induce a bound on the off-diagonal entries from the diagonal ones.

Our fit is motivated by recent experimental updates on key observables for the fit, such as the $W$-boson mass, the invisible width~\cite{Janot:2019oyi} of the $Z$-boson measured by LEP and the so-called Cabibbo anomaly, in which the sum of the first row of the CKM matrix is incompatible with 1 at the $2-3\sigma$ level. Additionally, we have taken the opportunity to improve the statistical analysis by taking into account correlations between observables and calibrating the test-statistic in order to gauge possible deviations from Wilks' theorem.

\section{Cases under study}
We separately consider the following heavy neutrino scenarios:
\begin{itemize}
    \item[$\bullet$] \textbf{2N-SS}: the minimal scenario compatible with oscillation data, showcasing two heavy neutrinos. This is a setup with little freedom in its parameter space, as the correct reproduction of the light neutrino mass matrix imposes strong correlations~\cite{Gavela:2009cd} between the entries of $\eta$. Furthermore, the Schwarz inequality is saturated ($|\eta_{\alpha\beta}|=\sqrt{\eta_{\alpha\alpha}\eta_{\beta\beta}}$), thus cLFV bounds must be combined in the fit with precision observables.
    \item[$\bullet$] \textbf{3N-SS}: the next-to-minimal scenario with three heavy neutrinos, which is the minimum number needed to have three massive light neutrinos. It can be shown that, in this setup, the correct reproduction of oscillation data induces correlations~\cite{Fernandez-Martinez:2015hxa} in the $\eta$-matrix. As in the previous scenario, the Schwarz inequality is also saturated, which again requires a combined fit of precision and cLFV observables.
    \item[$\bullet$] \textbf{G-SS}: the generic scenario in which more than three heavy neutrinos are present. Here all of the entries of $\eta$ are independent and are only subject to the constraints of Eq.~\ref{etapositive}. Since the Schwarz inequality is not saturated, precision and cLFV are fitted independently.
\end{itemize}

\section{Results and discussion}
\begin{table}
\centering
    \begin{tabular}{|c|c|c|c|c|c|}
    \hline 
    &\multicolumn{2}{c|}{}&\multicolumn{2}{c|}{}&\\[-2.5ex]
    \multirow{2}{*}{\bf95\%\,CL}&\multicolumn{2}{c|}{\bf 2N-SS} 
    &\multicolumn{2}{c|}{\bf 3N-SS}
    & \multirow{2}{*}{\bf G-SS}\\[0ex]  
    \cline{2-5} &&&&&\\[-2.5ex]
    & {\bf NO} & {\bf IO} & {\bf NO} & {\bf IO}  & \\
    \hline
    &&&&&\\[-2ex]
    $\eta_{ee}$ &    
    $9.4\cdot10^{-6}$&
    $5.5\cdot10^{-4}$&
    $1.3\cdot10^{-3}$&
    $1.4\cdot10^{-3}$&
    $[0.081, 1.4] \cdot10^{-3}$  \\[1.2ex]
    $\eta_{\mu\mu}$ &    
    $1.3\cdot10^{-4}$&
    $3.2\cdot10^{-5}$&
    $1.1\cdot10^{-5}$&
    $1.0\cdot10^{-5}$&
    $1.4 \cdot10^{-4}$  \\[1.2ex]
    $\eta_{\tau\tau}$  &    
    $2.1\cdot10^{-4}$&
    $4.5\cdot10^{-5}$&
    $1.0\cdot10^{-3}$&
    $8.1\cdot10^{-4}$&
    $8.9 \cdot10^{-4}$  \\[1.2ex]
    $\Tr{\eta}$ &    
    $2.9\cdot10^{-4}$&
    $6.0\cdot10^{-4}$&
    $1.9\cdot10^{-3}$&
    $1.5\cdot10^{-3}$&
    $2.1 \cdot10^{-3}$  \\[1.2ex]
    $\abs{\eta_{e\mu}}$ &    
    $1.2\cdot10^{-5}$&
    $1.3\cdot10^{-5}$&
    $1.2\cdot10^{-5}$&
    $1.2\cdot10^{-5}$&
    $1.2\cdot10^{-5}$\\[1.2ex]
    $\abs{\eta_{e\tau}}$ &    
    $2.2\cdot10^{-5}$&
    $1.4\cdot10^{-4}$&
    $9.0\cdot10^{-4}$&
    $8.0\cdot10^{-4}$&
    $8.8 \cdot10^{-4}$  \\[1.2ex]
    $\abs{\eta_{\mu\tau}}$&    
    $1.3\cdot10^{-4}$&
    $3.5\cdot10^{-5}$&
    $5.7\cdot10^{-5}$&
    $1.8\cdot10^{-5}$&
    $1.8 \cdot10^{-4}$  \\[1.2ex]
    
    \hline    
    \end{tabular}
    \caption{95$\%$\,CL upper bounds (or preferred intervals) for the elements of the non-unitarity $\eta$-matrix for the minimal setup with two heavy neutrinos (2N-SS), the next-to-minimal setup with three heavy neutrino (3N-SS) and the general case with an arbitrary number of heavy neutrinos (G-SS). For the first two scenarios, results are given for the two possible neutrino mass orderings: Normal Ordering (NO) and Inverted Ordering (IO). These results can be interpreted as bounds on the mixing of the heavy neutrinos with the active flavors.
    }
    \label{table:results}
\end{table}
The results of our global fit for the different cases under consideration are summarised in Table~\ref{table:results}. They are given as constraints on the different entries of the $\eta$-matrix, as well as the total deviation from unitarity, given by $\Tr{\eta}$. The test-statistics profiles from which the bounds are extracted can be found in the original paper~\cite{Blennow:2023mqx}. 

Table~\ref{table:results} shows that the bounds vary by orders of magnitude depending on which of the three cases is considered. First, the 2N-SS scenario is the most constrained one, showcasing bounds on the $\sim 10^{-5}-10^{-4}$ level. This can be understood in terms of the correlations imposed in $\eta$ from the correct reproduction of light neutrino masses and mixings. In particular, it can be shown~\cite{Gavela:2009cd} that this requires all entries of $\eta$ to be proportional to a common scale, which in turn means that all elements of $\eta$ must vanish simultaneously. This has important consequences for the fit, as the strong constraints on $\mu-e$ cLFV transitions forces $\abs{\eta_{e\mu}}$ to be very suppressed and, through the aforementioned correlation, this also forces the rest of elements of $\eta$ to be suppressed. Another notable feature of the 2N-SS bounds are the very different orders of magnitude obtained between the two possible light neutrino mass orderings: Normal Ordering (NO) and Inverted Ordering (IO). This is a consequence of the different flavor structures predicted by each of the orderings: while NO predicts a suppressed $\eta_{ee}$ with respect to $\eta_{\mu\mu}$ and $\eta_{\tau\tau}$, IO can allow for larger $\eta_{ee}$, which is slightly preferred by the data~\cite{Blennow:2023mqx}.

Secondly, the 3N-SS case has much looser bounds compared with the 2N-SS. This is a consequence of the fact that the correlations~\cite{Fernandez-Martinez:2015hxa} imposed within the $\eta$-matrix by oscillation data are looser when compared with those of the 2N-SS. Therefore, it is possible to evade the strong cLFV bounds while complying with the data. In particular, $\eta_{ee}$ and $\eta_{\tau\tau}$ are constrained at the $\sim10^{-3}$ level, which corresponds to the precision of the observables included in our fit. However, $\eta_{\mu\mu}$ receives a much stronger O$(10^{-5})$ bound from the saturation of the Schwarz inequality, which imposes a very strong bound on the product $\eta_{ee}\cdot\eta_{\mu\mu}$, and the small preference for a larger $\eta_{ee}$ that the data showcases.

Lastly, the G-SS case, which is the least restrictive of all the scenarios has, as expected, the weakest constraints. The bounds on the diagonal entries of $\eta$ are all at the $\sim10^{-3}$, as expected from the permille precision of our observables. Conversely, the situation of the off-diagonal entries is more interesting, as it is possible to constrain them directly from cLFV processes or indirectly from the Schwarz inequality, as previously discussed. We find that cLFV processes only put stronger constraints than the Schwarz inequality for $\abs{\eta_{e\mu}}$, since $\mu-e$ flavor transitions are very strongly constrained. The case of $\abs{\eta_{e\tau}}$ and $\abs{\eta_{\mu\tau}}$ is opposite, as cLFV transitions in the $\tau-\ell$ sector place much weaker bounds, and the induced bound from the Schwarz inequality dominates by more than an order of magnitude.

\section{Conclusions}
We have presented the results of an updated and improved global fit analysis of non-unitarity effects in the leptonic mixing matrix sourced by heavy neutrinos. These results can be interpreted as bounds on the mixing of the heavy neutrinos with the active flavors. These bounds apply and dominate for heavy neutrinos above the electroweak scale.

Our fit allows to simultaneously constrain all of the non-unitarity parameters at the $~10^{-5}-10^{-3}$. The constraints vary by orders of magnitude depending on which of the three studied scenarios is considered. Furthermore, the Schwarz inequality, arising from the positive-definiteness of the non-unitarity $\eta$-matrix, imposes stronger bounds on the off-diagonal entries mediating involving the $\tau$-flavor when compared with the bound directly extracted form the corresponding cLFV process. This implies that cLFV in the $\tau$-sector, if only sourced by heavy neutrinos, is constrained to lie far below the current experimental sensitivity.

\section*{Acknowledgments}
This project has received support from the European Union’s Horizon 2020 research and innovation programme under the Marie Skłodowska-Curie grant agreement No~860881-HIDDeN and No 101086085 - ASYMMETRY, and from the Spanish Research Agency (Agencia Estatal de Investigaci\'on) through the Grant IFT Centro de Excelencia Severo Ochoa No CEX2020-001007-S and Grant PID2019-108892RB-I00 funded by MCIN/AEI/10.13039/501100011033. DNT acknowledges support from the HPC-Hydra cluster at IFT. The work of DNT was supported by the
Spanish MIU through the National Program FPU (grant number FPU20/05333).

\section*{References}

\bibliography{biblio}

\end{document}